\documentclass[11pt]{article}
\usepackage{a4wide,latexsym,graphicx,epsfig,psfrag}
\usepackage{amsmath}
\usepackage{amsthm}
\usepackage{longtable}
\usepackage{array}
\usepackage{amssymb}
\usepackage{bbm}
\usepackage{cite}
\usepackage{accents}
\usepackage{pgfsys}
\usepackage{dsfont}
\newcommand{\nn}{\nonumber \\}
\newcommand{\no}{\nonumber}

\newcommand{\vp}{\varphi}

\newcommand{\lra}{\longrightarrow}

\newcommand\lsim{\mathrel{\rlap{\lower4pt\hbox{\hskip1pt$\sim$}}
    \raise1pt\hbox{$<$}}}
\newcommand\gsim{\mathrel{\rlap{\lower4pt\hbox{\hskip1pt$\sim$}}
    \raise1pt\hbox{$>$}}}
\newcommand{\as}{\accentset}
\newcommand{\cM}{{\cal M}}
\begin{document}
\parskip=3pt plus 1pt

\begin{titlepage}
\vskip 1cm
\begin{flushright}
{\sf  UWThPh-2013-28 \\
  } 
\end{flushright}

\setcounter{footnote}{0}
\renewcommand{\thefootnote}{\fnsymbol{footnote}}

\vspace*{1.5cm}
\begin{center}
{\Large\bf Approximating chiral SU(3) amplitudes } 
\\[25mm]

{\normalsize\bf G. Ecker$^{1}$, P. Masjuan$^{2}$ and
  H. Neufeld$^{1}$}\\[1.2cm]
\end{center}   
\begin{flushleft} 
${}^{1)}$ University of Vienna, Faculty of Physics,
Boltzmanngasse 5, A-1090 Wien, Austria \\[10pt]
${}^{2)}$ Institut f\"ur Kernphysik, Johannes Gutenberg-Universit\"at,
D-55099 Mainz, Germany
\end{flushleft}

\vspace*{2cm}

\begin{abstract}
\noindent
We construct large-$N_c$ motivated approximate chiral $SU(3)$
amplitudes of next-to-next-to-leading order. The amplitudes are
independent of the renormalization scale. Fitting lattice data with
those amplitudes allows for the extraction of chiral coupling
constants with the correct scale dependence. The differences between
approximate and full amplitudes are required to be at most of the
order of N$^3$LO contributions numerically. Applying the approximate
expressions to recent lattice data for meson decay constants, we determine
several chiral couplings with good precision. In particular, we obtain
a value for $F_0$, the meson decay constant in the chiral $SU(3)$
limit, that is more precise than all presently available determinations.
\end{abstract}

\vfill

\setcounter{footnote}{0}
\renewcommand{\thefootnote}{\arabic{footnote}} 

\end{titlepage}

%\tableofcontents

\newpage

\section{Introduction} 
\label{sec:intro}
\renewcommand{\theequation}{\arabic{section}.\arabic{equation}}
\setcounter{equation}{0}
Hadronic processes at low energies cannot be treated with
perturbative QCD. The main protagonists in this field, lattice QCD 
and chiral perturbation theory (CHPT), have mutually benefited 
from a cooperation started several years ago. The emphasis of
this cooperation has shifted in recent years. Although extrapolation
to the physical quark (and hadron) masses and finite-volume
corrections, both accessible in CHPT, are still useful for lattice
simulations, improved computing facilities and
lattice algorithms allow
for simulations with ever smaller quark masses and larger volumes. 
On the other hand, the input of lattice QCD for CHPT has become more
important over the years to determine the coupling constants of chiral
Lagrangians, the so-called low-energy constants (LECs). This input is
especially welcome for LECs modulating quark mass terms:
unlike in phenomenological analyses, quark (and hadron) masses
can be tuned on the lattice. 

While this program has been very successful for chiral $SU(2)$, the
situation is less satisfactory for $SU(3)$ 
\cite{Aoki:2013ldr}. In the latter case, the natural expansion
parameter (in the meson sector) is $M_K^2/16 \pi^2 F_\pi^2 \simeq
0.2$. To match the precision that lattice studies can attain nowadays,
it is therefore mandatory to include NNLO contributions in
CHPT. Although NNLO amplitudes are available for most quantities of
interest for lattice simulations \cite{Bijnens:2006zp}, there has been
a certain reluctance in the lattice community to make full use of
those amplitudes for two reasons mainly: for chiral $SU(3)$, NNLO
amplitudes are usually quite involved and they are mostly available in
numerical form only. 

In this paper, we resume our proposal 
\cite{Ecker:2010nc} for large-$N_c$ motivated approximations of NNLO
amplitudes that contain one-loop functions only. Besides
recapitulating  the main features of those analytic
approximations, the following issues will be discussed.
\begin{itemize}
\item We set up numerical criteria for the amplitudes to qualify as
  acceptable approximations. Those criteria can be checked by
  comparing with available numerical results making use of the full
  NNLO amplitudes for some given sets of meson masses.
\item The proposed approximation includes all terms leading and
  next-to-leading order in large $N_c$. In addition, it contains all
  chiral logs, independently of the large-$N_c$ counting. In order to meet
  the numerical criteria just mentioned, it may sometimes be useful to
  go beyond the strict large-$N_c$ counting by including also products
  of one-loop functions occurring in two-loop diagrams.
\item In addition to the ratio $F_K/F_\pi$ of meson decay constants
  investigated in 
  Ref.~\cite{Ecker:2010nc}, here we also study the pion decay constant
  $F_\pi$ itself. By confronting our approximation with 
  lattice data, we demonstrate the possibilities to extract information
  on both NLO and NNLO LECs. While the NNLO LECs have the expected large
  uncertainties, the NLO LECs can be determined quite well. Our
  numerical fits of lattice data are not intended to compete with
  actual lattice results for obvious reasons. Instead, we hope 
  to encourage lattice groups to use NNLO amplitudes 
  that are much simpler than the full amplitudes and
  yet offer considerably more insight than, e.g., polynomial
  fits. These amplitudes can also be considered as relatively
  simple tools to study convergence issues of chiral $SU(3)$ with
  lattice data. 
\end{itemize}     

In Sec.~\ref{sec:anapp} we review the structure and the salient
features of the approximate form of NNLO amplitudes for chiral $SU(3)$
proposed in Ref.~\cite{Ecker:2010nc}. In addition to setting up a
criterion for deciding whether the approximation is acceptable for a
given observable, we also suggest a possible modification of the
original version. Both approximations are applied to an analysis of
lattice data for the ratio $F_K/F_\pi$ to extract NLO and NNLO
LECs. We study in detail the dependence of the approximations on a
scale parameter $M$ that mimics the neglected two-loop
contributions. The extracted LECs are then used in Sec.~\ref{sec:fpi}
to analyse $F_\pi$ in chiral $SU(3)$. It turns out that $F_\pi$ is
well suited for determining the leading-order LEC $F_0$, the meson decay
constant in the chiral $SU(3)$ limit. We demonstrate why the NLO LEC
$L_4$ is usually strongly anti-correlated with $F_0$ in
phenomenological analyses. We also discuss the constraints on $F_0$
coming from a comparison with chiral $SU(2)$. Sec.~\ref{sec:kl3}
contains a few remarks on the kaon semileptonic vector form factor at
$t=0$. In App.~\ref{app:gfp6} we rederive the generating functional of
Green functions at NNLO \cite{Bijnens:1999hw} in a form suitable for
our analytic approximations. Explicit approximate expressions for
$F_K/F_\pi$ and $F_\pi$, which are the basis for the analysis in 
previous sections, are presented in Apps.~\ref{app:fkfpi} and
\ref{app:fpif0}, respectively. 

\section{Analytic approximations of NNLO amplitudes} 
\label{sec:anapp}
\begin{figure}[!t]
\centerline{\epsfig{file=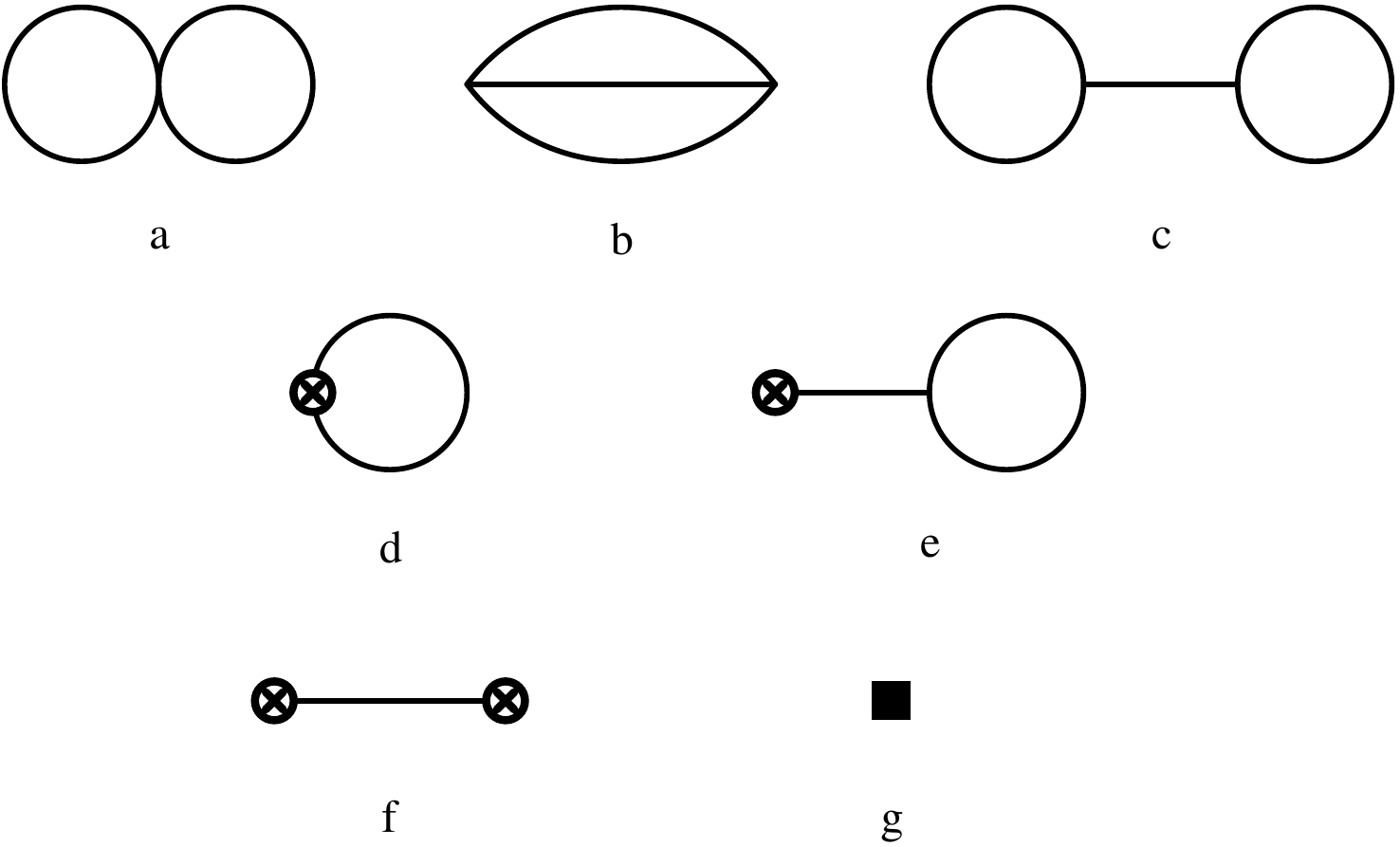,height=7cm}}
\caption{Skeleton diagrams for the generating functional
  $Z_6$ of  $O(p^6)$. Simple dots, crossed circles, black box denote
  vertices from LO, NLO, NNLO Lagrangians,
  respectively. Propagators and vertices carry the full tree structure 
  associated with the lowest-order Lagrangian. }
\label{fig:p6diag}
\end{figure}

\renewcommand{\theequation}{\arabic{section}.\arabic{equation}}
\setcounter{equation}{0}
CHPT can be formulated in terms of the generating functional of Green
functions $Z[j]$ \cite{Gasser:1983yg,Gasser:1984gg}. The NNLO
functional $Z_6$ of $O(p^6)$ is a sum of various contributions shown
in Fig.~\ref{fig:p6diag}. In App.~\ref{app:gfp6}, we derive an
explicit representation of $Z_6$ based on the work of
Ref.~\cite{Bijnens:1999hw}.  
In Ref.~\cite{Ecker:2010nc}, we proposed an analytic approximation for 
$Z_6$ of the following form:
\begin{eqnarray} 
Z_6^{I} &=&  \int \!\! d^4x \left\{ \left[C_a^r(\mu) +
\displaystyle\frac{1}{4 F_0^2} \left(4\, \Gamma_a^{(1)} \,L(\mu)  -
\Gamma_a^{(2)} \,L(\mu)^2 + 2\, \Gamma_a^{(L)}(\mu) L(\mu)
\right)\right] O_a(x) \right. \no \\[.2cm]
&& + \left. \displaystyle\frac{1}{(4\pi)^2} \left[L_i^r(\mu) - 
  \displaystyle\frac{\Gamma_i }{2} L(\mu) \right] H_i(x;M) \right\} \no 
\\[.2cm]
&+&  \int \!\! d^4x\,d^4y \left\{\left(L_i^r(\mu) - 
  \displaystyle\frac{\Gamma_i }{2} L(\mu) \right)
  P_{i,\alpha}(x) \,G_{\alpha,\beta}(x,y)  
  \left(L_j^r(\mu) - 
  \displaystyle\frac{\Gamma_j }{2} L(\mu) \right) P_{j,\beta}(y)\right.
  \no \\[.1cm] 
&& + \left. 2\,\left(L_i^r(\mu) - 
  \displaystyle\frac{\Gamma_i }{2} L(\mu) \right)
  P_{i,\alpha}(x) \,G_{\alpha,\beta}(x,y)\, F_\beta(y;M)\right\} ~.
\label{eq:approxI} 
\end{eqnarray}   
The monomials $O_a(x) ~(a=1,\dots,94)$ define the chiral Lagrangian of
$O(p^6)$ \cite{Bijnens:1999sh} with associated renormalized LECs
$C_a^r(\mu)$ and  
the $L_i^r(\mu) ~(i=1,\dots,10)$ are renormalized LECs of $O(p^4)$ with
associated beta functions $\Gamma_i$ \cite{Gasser:1984gg}. The
coefficients $ \Gamma_a^{(1)}$, $\Gamma_a^{(2)}$ and $\Gamma_a^{(L)}$
are listed in Ref.~\cite{Bijnens:1999hw}. Repeated indices are to be
summed over. $F_0$ is the meson decay constant in the chiral $SU(3)$
limit. The chiral log
\begin{equation} 
L(\mu) = \displaystyle\frac{1}{(4\pi)^2} \ln{M^2/\mu^2} 
\label{eq:clog}
\end{equation}
involves an arbitrary scale $M$. This scale is introduced in the
complete functional  $Z_6$ in Eq.~(\ref{eq:Ztotal}) to make the scale
dependence explicit: only $C_a^r(\mu)$, $L_i^r(\mu)$ and $L(\mu)$ in 
(\ref{eq:approxI}) depend on
the renormalization scale $\mu$. The various functionals in 
Eq.~(\ref{eq:approxI}) are defined in App.~\ref{app:gfp6}. The
approximation consists in dropping in $Z_6$ the irreducible two-loop
contributions represented by the functional $K(x;M)$ (diagrams a,b in
Fig.~\ref{fig:p6diag}) and the terms bilinear in $F_\alpha(x;M)$
(diagram c in Fig.~\ref{fig:p6diag}), except for
single and double logs.

The procedure how to actually calculate an amplitude corresponding to
Eq.~(\ref{eq:approxI}) was described in Ref.~ \cite{Ecker:2010nc}. In
many cases, the relevant amplitudes can be extracted from available
calculations of $O(p^6)$ \cite{Bijnens:2006zp}.   

Approximation I defined by Eq.~(\ref{eq:approxI}) has the following
properties \cite{Ecker:2010nc}:
\begin{itemize} 
\item All chiral logs are included.
\item The functional $Z_6^{I}$ is independent of the renormalization
  scale $\mu$. Unlike the double-log approximation\cite{Bijnens:1998yu}, 
  it therefore
  allows for the extraction of LECs with the correct scale dependence.
\item In addition to single and double logs, the residual dependence
  on the scale $M$ is the only other vestige of the two-loop
  part.
\item In dropping the genuine two-loop contributions, Approximation I
  respects the large-$N_c$ hierarchy of $O(p^6)$ contributions:
  \\[.3cm] 
  \hspace*{2cm} $C_a, L_i L_j \quad\lra\quad L_i \times
  $\,loop $\quad\lra\quad$  two-loop contributions \\[-.1cm] 
\item Only tree and one-loop amplitudes need to be calculated.
\end{itemize} 

The question still remains to be answered how reliable this
approximation is. We shall adopt the following criterion. For an
$SU(3)$ quantity normalized to one at lowest order, successive terms
in the chiral expansion usually
show the following generic behaviour:  
\begin{center} 
\begin{tabular}{ccc}
 & & \\[-.1cm] 
\hspace*{.5cm} $O(p^4)$ \hspace*{.5cm} & $O(p^6)$ & $O(p^8)$  \\[.2cm] 
$\lsim 0.3$ & \hspace*{.2cm} $\lsim 0.3^2 = 0.09$ \hspace*{.2cm} &
 \hspace*{.1cm} $\lsim 0.3^3 = 0.027$  \hspace*{.1cm} \\[-.1cm] 
 & & 
\end{tabular}  
\end{center}    
This suggests as a criterion for an acceptable NNLO approximation that      
the accuracy should not be worse than 3 $\%$, the typical size of 
contributions of $O(p^8)$.  As the following examples will show, the
quality of the approximation depends on the scale $M$, which
parametrizes the two-loop contributions not contained in
(\ref{eq:approxI}). Although the acceptable range will depend on the
quantity under consideration, experience with the double-log
approximation \cite{Bijnens:1998yu} in chiral $SU(3)$ suggests that
$M$ is of the order of $M_K$. 
 
Approximation I is motivated by large $N_c$, but in some cases the
accuracy may be improved by including in the approximate functional
(\ref{eq:approxI}) also products of one-loop amplitudes (diagrams a,c in
Fig.~\ref{fig:p6diag}, subleading in $1/N_c$), which also have a simple
analytic form. 
We call this extension Approximation II. In contrast to
Approximation I, this extension is not uniquely
defined\footnote{Hans Bijnens, private communication} because it
depends on the representation of the matrix field $U$. In the standard
representation used, e.g., also in Refs.~\cite{Bijnens:1999hw,Amoros:1999dp},  
it amounts to omitting (except for chiral logs) the sunset
diagram b from the full functional $Z_6$ in Eq.~(\ref{eq:Ztotal}).

\section{$\mathbf{F_K/F_\pi}$ and the low-energy constant
  $\mathbf{L_5}$}  
\label{sec:fkfpi}
\renewcommand{\theequation}{\arabic{section}.\arabic{equation}}
\setcounter{equation}{0}
The ratio of pseudoscalar decay constants $F_K/F_\pi$ appears well
suited for our analytic approximations. The chiral expansion
of $F_K/F_\pi -1$ is shown for physical meson masses in Table
\ref{tab:physmass}. The separately scale-dependent contributions of
$O(p^6)$ are given for the usual renormalization scale $\mu=770$ MeV.
The entries for ``numerical results'' were provided by Bernard and
Passemar \cite{Bernard:2009ds}. 
\begin{center}   
\begin{table}[!h]
\begin{center} 
\begin{tabular}{lcccc}
\hline
  & & & & \\
  & $O(p^4)$  &  &   $O(p^6)$ &   \\[.2cm] 
  & &\hspace*{.2cm} 2-loop \hspace*{.2cm} & \hspace*{.2cm} $L_i
\times$ loop \hspace*{.2cm} & \hspace*{.2cm} tree  \hspace*{.2cm} \\[.2cm]
numerical results \cite{Amoros:1999dp,Bernard:2009ds} & \hspace*{.2cm}
  0.14 \hspace*{.2cm}  & 0.002 & 0.051 & 
  0.008 \\[.1cm] 
Approximation I ($M=M_K$) &      & - 0.030 & & \\[.1cm]
Approximation II ($M=M_K$) &      & - 0.011 & & \\[.1cm]  
\hline
\end{tabular}
\caption{Chiral expansion of $F_K/F_\pi -1$. The separate contributions of
$O(p^6)$ are listed for $\mu=770$ MeV.} 
\label{tab:physmass}
\end{center} 
\end{table}
\end{center}
\begin{center} 
%\vspace*{-4cm} 
\begin{figure}[!h]
\setlength{\unitlength}{1cm}
\begin{center} 
\begin{picture}(20,4)
\put(0.1,0.03){\makebox(9.0,8.0)[lb]{
\leavevmode 
\includegraphics[width=6.5cm]{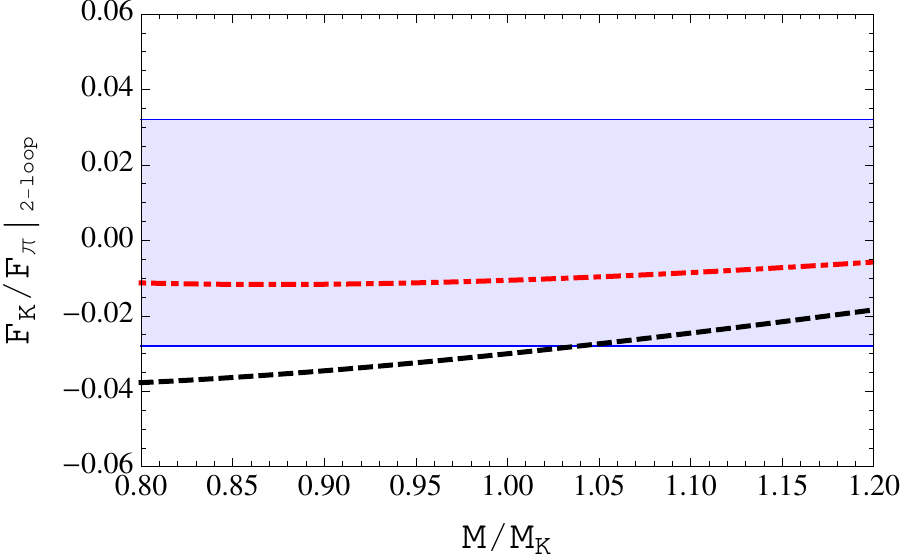}}}
\put(8.5,0.03){\makebox(9.0,8.0)[lb]
{\leavevmode 
\includegraphics[width=6.5cm]{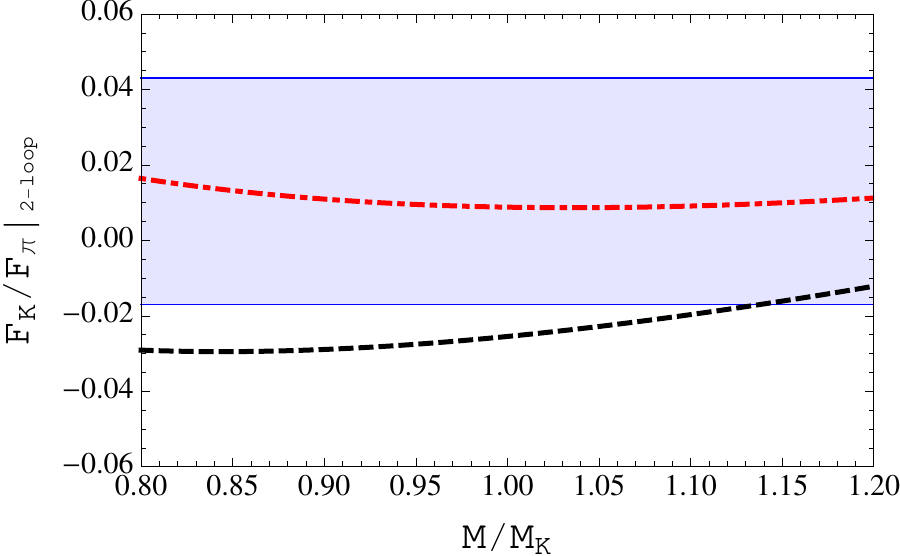}}}
\end{picture}
\caption{$M$-dependence of Approximations I, II for the two-loop
  contributions to $F_K/F_\pi$ at $\mu=770$ MeV. The blue bands denote
  the $\pm 3 \%$ regions around the actual values provided by
  Bernard and Passemar \cite{Bernard:2009ds}. The dashed black curves
  correspond to 
  Approximation I, the dash-dotted red curves to Approximation
  II. The left panel describes the situation for physical meson
  masses, the right panel for masses $M_\pi=416$ MeV, $M_K=604$ MeV.}  
\label{fig:MFKFpi}
\end{center} 
\end{figure}
\end{center} 
As shown in Table \ref{tab:physmass}, Approximation I barely meets our
criterion of acceptability put forward at the end of the last section,
while Approximation II does much better. To
investigate also the dependence on the scale $M$, we display in
Fig.~\ref{fig:MFKFpi} the variation 
with $M$ for both versions I and II and for two sets of meson
masses. As demonstrated in Fig.~\ref{fig:MFKFpi}, 
there is little dependence on $M$ in the vicinity of $M=M_K$.
Nevertheless, we will account for this variation in the final errors
for the preferred Approximation II.  

The explicit expression for Approximation II of $F_K/F_\pi$ is given
in App.~\ref{app:fkfpi} where all masses are lowest-order masses of
$O(p^2)$. Since we work to $O(p^6)$ the masses in
$R_4$ must be expressed in terms of the lattice masses to
$O(p^4)$ \cite{Gasser:1984gg}. The chiral limit value $F_0$ is
expressed in terms of the experimental value $F_\pi=92.2$ MeV and
physical meson masses, using again the $O(p^4)$ relation. In $R_6$
and $R_6^{\rm ext}$, $F_0$ and the meson masses can be replaced by
$F_\pi$ and lattice masses, respectively.

We now repeat the analysis of $F_K/F_\pi$ performed in
Ref.~\cite{Ecker:2010nc} with Approximation II. We recall that at
$O(p^4)$ only the LEC $L_5$ enters. At $O(p^6)$, two
combinations of NNLO LECs appear: $C_{14} + C_{15}$ and $C_{15} + 2
C_{17}$. At this order, also some of the $L_i$ enter, but only the
term with $L_5^2$ is leading in $1/N_c$. In the spirit of large
$N_c$, we therefore extract as in Ref.~\cite{Ecker:2010nc} $L_5$, 
$C_{14} + C_{15}$ and $C_{15} + 2 C_{17}$ from a fit of the lattice data
of the BMW Collaboration \cite{Durr:2010hr}, using for the remaining
$L_i$ (appearing only at $O(p^6)$, subleading in  $1/N_c$) the
values of fit 10 of Ref.~\cite{Amoros:2001cp}.  
The results are displayed in Table \ref{tab:results}. 
\begin{center} 
\begin{table}[ht]
\begin{center} 
\begin{tabular}{lcccc}
\hline
 & & & & \\[-.1cm] 
  & $F_K/F_\pi$  & $10^3 L_5^r$
  & $10^3 (C_{14}^r + C_{15}^r)$
  & $10^3 (C_{15}^r + 2 C_{17}^r)$ \\[.2cm]
\hline  \\[-.1cm]  
App. I ($M=M_K$) &  $1.198(5)$    & $0.76(8)$ & $0.37(7)$ & $1.29(14)$ 
 \\[.1cm] 
App. II  & $1.200(5)$    & $0.75(8)$ & $0.20(8)$ & $0.71(22)$ 
 \\[.1cm]
BMW \cite{Durr:2010hr} & \hspace*{.2cm} $1.192(7)_{\rm stat}(6)_{\rm syst}$
  \hspace*{.2cm}  & & & \\[.1cm] 
\hline
\end{tabular}
\caption{Fit results for $F_K/F_\pi$ and LECs for Approximations I
  (statistical errors only) and II. The renormalization scale is
  $\mu=770$ MeV for all LECs. The LECs $C_a^r$ have dimension 
  ${\rm GeV}^{-2}$.} 
\label{tab:results}
\end{center} 
\end{table}
\end{center} 
The fitted values of $F_K/F_\pi$ agree with the detailed analysis of
Ref.~\cite{Durr:2010hr}. For both $F_K/F_\pi$ and $L_5$, there is
practically no difference between the two approximations but the LECs
of $O(p^6)$ show a bigger spread. For Approximation II, we have added
the uncertainty due to varying $M$ in the range $0.9 \le M/M_K \le
1.1$ in quadrature to the 
statistical lattice errors. The effect of this variation is small,
for $F_K/F_\pi$ and $L_5$ in fact negligible. Since $C_{15}$
is subleading in $1/N_c$ the fit determines essentially $C_{14}$ and
$C_{17}$ \cite{Bernard:2009ds}. Although the values depend of 
course on the input for the $L_i$, the results in Table
\ref{tab:results} suggest that both $C_{14}^r$ and $C_{17}^r$ are
positive and smaller than $10^{-3} ~{\rm GeV}^{-2}$, always taken at
$\mu=0.77$ GeV. We will use these fit results with Approximation II
for $L_5^r$, $C_{14}^r$ and $C_{17}^r$ in the analysis of $F_\pi/F_0$
in the following section.

The fit also demonstrates very clearly that NNLO terms are 
essential. While the NNLO fit (Approximation II) is well behaved
($\chi^2$/dof = 1.2, statistical errors only), the NLO fit with the
single parameter $L_5$ is unacceptable ($\chi^2$/dof =
4). Analysing present-day lattice data with NLO chiral $SU(3)$
expressions does not make sense.

\section{ $\mathbf{F_\pi}$ and the low-energy constants  $\mathbf{F_0,L_4}$} 
\label{sec:fpi}
\renewcommand{\theequation}{\arabic{section}.\arabic{equation}}
\setcounter{equation}{0}
The meson decay constant in the chiral limit is a LEC of the
lowest-order chiral Lagrangian. In
the case of chiral $SU(2)$, $F =  \lim_{m_u,m_d \to 0} F_\pi$ is well
known, mainly from a combined analysis of lattice data with $N_f=2$
active flavours by the FLAG Collaboration \cite{Aoki:2013ldr}: 
\begin{equation} 
F = (85.9 \pm 0.6) ~{\rm MeV}~.
\label{eq:F}
\end{equation}
The situation is quite different in the $SU(3)$ case. The lattice
results for $F_0 =  \lim_{m_u,m_d,m_s \to 0} F_\pi$ cover a much wider
range, from about 66 MeV to 84 MeV
\cite{Aoki:2013ldr}. A similar range is covered in the
phenomenological fits of Bijnens and Jemos \cite{Bijnens:2011tb}. 

The low-energy expansion in chiral $SU(3)$ is characterized by the
ratio $p^2/(4\pi F_0)^2$ where $p$ stands for a generic meson momentum
or mass. $F_0$ thus sets the scale for the chiral expansion. In
practice, $F_0$ is usually traded for $F_\pi$ at successive orders of
the chiral expansion. Nevertheless, $F_0$ affects the
``convergence'' of the chiral expansion: a smaller $F_0$ tends to
produce bigger fluctuations at higher orders. 

Why has it been so difficult both for lattice and phenomenological
studies to determine $F_0$? One clue is the apparent
anti-correlation with the NLO LEC $L_4$ in the fits of
Ref.~\cite{Bijnens:2011tb}: the bigger $F_0$, the smaller
$L_4^r(M_\rho)$, and vice versa. The large-$N_c$ suppression of $L_4$
is not manifest in the fits with small $F_0$. 

This anti-correlation can be understood to some extent from the
structure of the chiral $SU(3)$ Lagrangian up to and including NLO:  
\noindent
\begin{eqnarray}
{\cal L}_{p^2}(2) + {\cal L}_{p^4}(10) &=&
\displaystyle\frac{F_0^2}{4} \langle 
D_\mu U D^\mu U^\dagger +  \chi U^\dagger + \chi^\dagger  U \rangle 
+ L_4 \langle D_\mu U D^\mu U^\dagger\rangle 
\langle \chi U^\dagger + \chi^\dagger  U \rangle + \dots  \nn
&=& \displaystyle\frac{1}{4} \langle D_\mu U
D^\mu U^\dagger \rangle \left[ F_0^2 + 8 L_4 \left(2 \as{\circ}{M}^2_K +
  \as{\circ}{M}^2_{\pi} \right)\right] + \dots~ 
\end{eqnarray} 
The unitary matrix field $U$ is parametrized by the meson fields,
$\chi = 2B_0 \cM_q$ ($B_0 \sim$ quark condensate, $\cM_q$ is the quark
mass matrix), $\langle \dots \rangle$ stands for the $SU(3)$ flavour
trace and $\as{\circ}{M}_P$ denotes the lowest-order meson masses. The
dots refer to the remainder of the NLO Lagrangian in the first
line and to terms of higher order in the meson fields in the second
line. Therefore, a LO tree-level contribution is always
accompanied by an $L_4$ contribution in the combination 
\begin{equation} 
F(\mu)^2:=F_0^2 + 8 L_4^r(\mu) \left(2 \as{\circ}{M}^2_K +
\as{\circ}{M}^2_{\pi}\right)~.
\label{eq:Fmu}
\end{equation} 
Of course, there will in general be additional contributions involving
$L_4$ at NLO, especially in higher-point functions (e.g., in meson
meson scattering). Nevertheless, the observed anti-correlation between
$F_0$ and $L_4$ is clearly related to the structure of the chiral
Lagrangian. Note that $F^2_{\pi}/16 M_K^2 = 2 \times 10^{-3}$ is the
typical size of a NLO LEC. Although of different chiral order, the two
terms in $F(\mu)^2$ could a priori be of the same order of magnitude.

\begin{center} 
\begin{figure}[!ht]
\begin{center}  
\leavevmode
\includegraphics[width=10cm]{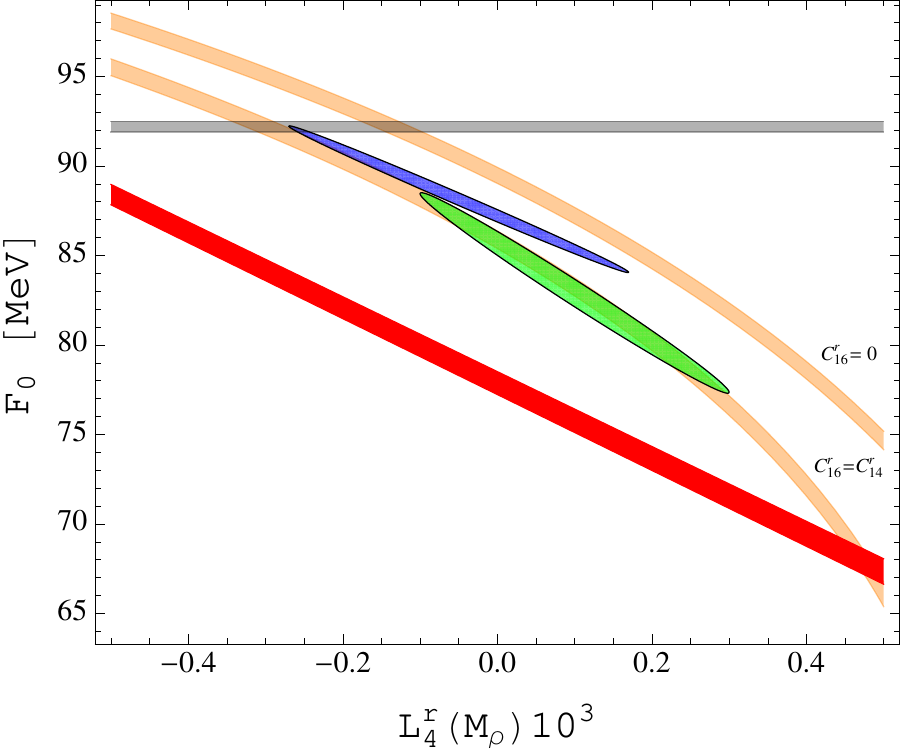} 
\end{center} 
\caption{Fitted values of $F_0$, $L_4$ using RBC/UKQCD data
  \cite{Aoki:2010dy,Arthur:2012opa} with $M_\pi < 350$ MeV, with (blue
  ellipse) and without (green ellipse) including $F_\pi^{\rm
  phys}$. The red band results from the  
  comparison of $F_\pi$ between $SU(2)$ and $SU(3)$ as expressed by
  Eq.~(\ref{eq:F0F}), taking $F  = (85.9 \pm 0.6) ~{\rm MeV}$ 
  \cite{Aoki:2013ldr}. 
  The relation between $F_0$ and $F$ to $O(p^6)$
  \cite{Gasser:2007sg} leads to the orange bands for two values of
  $C^r_{16}(M_\rho)$.
  The horizontal grey band denotes
  $F_\pi = (92.2 \pm 0.3)$ MeV.}
\label{fig:F0FL4_34}
\end{figure}
\end{center}

Independent information on $F_0$ comes from comparing the $SU(2)$ and
$SU(3)$ expressions for $F_\pi$.  To $O(p^4)$ in chiral $SU(2)$,
$F_\pi$ is given by \cite{Gasser:1983yg} 
\begin{equation} 
F_\pi = F + F^{-1} \left[M_\pi^2 \,l_4^r(\mu) + \overline{A}(M_\pi,\mu)
\right]
\label{eq:su2}
\end{equation}
where $l_4$ is a NLO $SU(2)$ LEC and $\overline{A}(M_\pi,\mu)$ is a
one-loop function defined in Eq.~(\ref{eq:1loop}).
Expressing $l_4^r(\mu)$  in terms of $L_4^r(\mu)$, $L_5^r(\mu)$
and a kaon loop contribution \cite{Gasser:1984gg} and equating
Eq.~(\ref{eq:su2}) with the $SU(3)$ result for $F_\pi$, one arrives at
the following relation:
\begin{eqnarray}  
F_0 &=& F - F^{-1}\left\{\left(2 M_K^2 - M_\pi^2\right)\left(4 L_4^r(\mu) +
\displaystyle\frac{1}{64 \pi^2}\ln{\displaystyle\frac{\mu^2}{M_K^2}}\right)
+ \displaystyle\frac{M_\pi^2}{64 \pi^2}  \right\} + O(p^6) ~.
\label{eq:F0F}
\end{eqnarray} 
To $O(p^6)$, the relation between $F_0$ and $F$ was derived by Gasser
et al. \cite{Gasser:2007sg}. It depends on LECs of both NLO and NNLO. In
Fig.~\ref{fig:F0FL4_34}, both $O(p^4)$ and $O(p^6)$ relations will be
displayed. Of course, in order to plot $F_0$ as a function of $L_4$ to
$O(p^6)$, some assumptions about NLO and NNLO LECs are needed. 

$SU(3)$ lattice data for $F_\pi$ seem well suited for a determination
of $F_0$ and $L_4$ although the emphasis in most lattice studies has
been to determine $F_\pi$ itself. As for $F_K/F_\pi$, 
the use of CHPT to NNLO, $O(p^6)$ \cite{Amoros:1999dp}, is essential
for a quantitative analysis. 

In the following, we are going to apply Approximation I for the analysis
of $F_\pi$. It turns out that, unlike for $F_K/F_\pi$, Approximation I
agrees better with the numerical results of Ref.~\cite{Bernard:2009ds}
than Approximation II. The explicit representation for $F_\pi$ is
given in App.~\ref{app:fpif0}. The lowest-order masses appearing in
the terms of $O(p^4)$ must again be expressed in terms of lattice
masses. Unlike in the previous section, we leave $F_0$ in
Eq.~(\ref{eq:fpiI}) untouched.

\begin{center} 
%\vspace*{-4cm} 
\begin{figure}[!h]
\setlength{\unitlength}{1cm}
\begin{center} 
\begin{picture}(20,4)
\put(0.1,0.03){\makebox(9.0,8.0)[lb]{
\leavevmode 
\includegraphics[width=6.5cm]{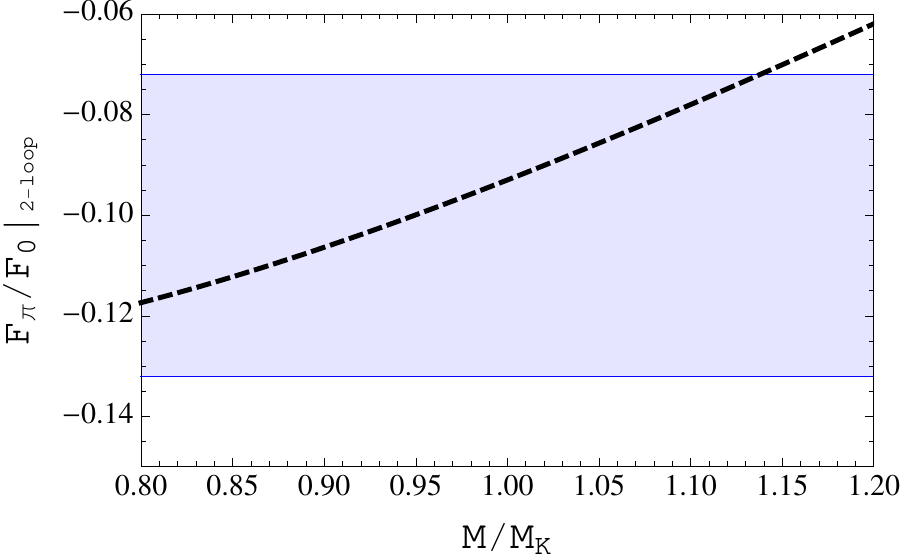}}}
\put(8.5,0.03){\makebox(9.0,8.0)[lb]
{\leavevmode 
\includegraphics[width=6.5cm]{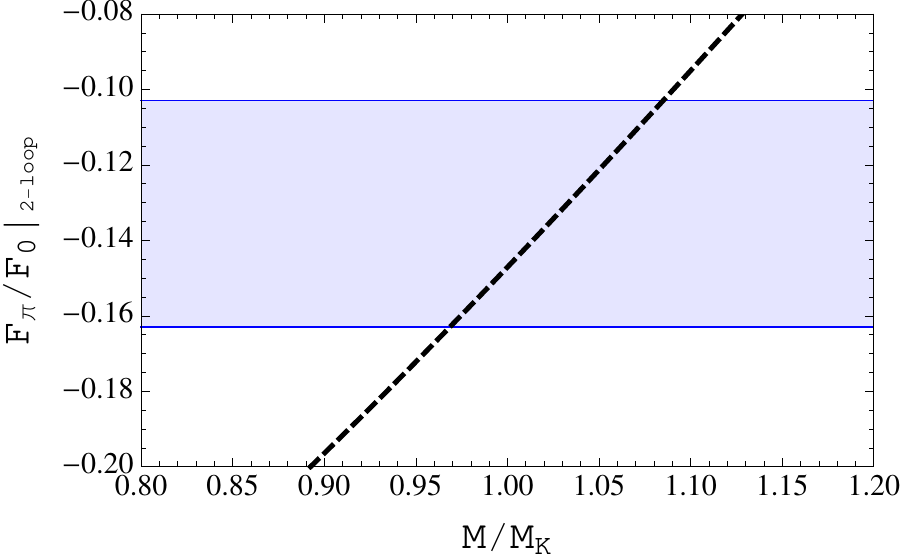}}}
\end{picture}
\caption{$M$-dependence of Approximation I for the two-loop
  contributions to $F_\pi$ at $\mu=770$ MeV (dashed black
  curves). The blue bands denote the $\pm 3 \%$ regions around the
  actual values provided by Bernard and Passemar \cite{Bernard:2009ds}. 
  The left panel describes the situation for physical meson
  masses, the right panel for masses $M_\pi=416$ MeV, $M_K=604$ MeV.}  
\label{fig:MFpi}
\end{center} 
\end{figure}
\end{center}

Again in contrast to the ratio
$F_K/F_\pi$, the dependence on the mass parameter $M$ is more
pronounced in this case, especially for larger meson masses (see
Fig.~\ref{fig:MFpi}). To satisfy the re\-quirement that our
approximation should stay within $\pm 3 \%$ of the exact numerical
results \cite{Bernard:2009ds}, we are going to vary $M$ in the range
$0.97 \le M/M_K \le 1.09$. 
In addition to $F_0$ and $L_4$, the only other LEC appearing to
$O(p^4)$ in Eq.~(\ref{eq:fpiI}) is $L_5$. On the basis of the analysis
of $F_K/F_\pi$ in Sec.~\ref{sec:fkfpi}, we will use
$L_5^r= (0.75 \pm 0.10) \cdot 10^{-3}$. At $O(p^6)$, the
following NNLO LECs enter: $C_{14}$, $C_{15}$, $C_{16}$ and $C_{17}$,
but only  $C_{14}$ and $C_{17}$ are leading in $1/N_c$. In the spirit
of large $N_c$, we therefore use the values for $C_{14}$ and $C_{17}$
obtained in the previous section, neglecting at the same time $C_{15}$
and $C_{16}$. However, we assign a 100 $\%$ uncertainty to both
$C_{14}$ and $C_{17}$. 
Anticipating the dependence of the relation
between $F_0$ and $F$ at $O(p^6)$ on $C_{16}$ \cite{Gasser:2007sg} in
Fig.~\ref{fig:F0FL4_34}, we include for consistency the
uncertainty due to varying $C^r_{16}(M_\rho)$ between 
$\pm C^r_{14}(M_\rho)$. 
At $O(p^6)$, some more of the NLO LECs $L_i$
enter. For definiteness, we use again fit 10 of
Ref.~\cite{Amoros:2001cp} for those LECs. However, any other set of
values for the $L_i$ from Refs.~\cite{Amoros:2001cp,Bijnens:2011tb}
consistent with large $N_c$, in particular with a small
$L_4^r(M_\rho)$, leads to very similar results.    

We confront the expression (\ref{eq:fpiI}) for $F_\pi$ with
lattice data from the RBC/UKQCD Collaboration
\cite{Aoki:2010dy,Arthur:2012opa}. In our main fit we only consider
(five) unitary lattice points with $M_\pi < 350$ MeV. In this case,
$F_\pi$ for physical meson masses emerges as a fit result but the
fitted value is lower than the experimental value. Another 
alternative is therefore to use in addition to the lattice points also
the experimental value $F_\pi = (92.2 \pm 0.3)$ MeV as input where we 
have doubled the error assigned by the Particle Data Group
\cite{Beringer:1900zz}.   
The extracted values of $F_0$ and $L_4^r(M_\rho)$ are shown in
Fig.~\ref{fig:F0FL4_34}. For the case where $F_\pi^{\rm phys}$ is
included (blue ellipse), the explicit fit results are: 
\begin{eqnarray}
F_0 &=& (88.1 \pm 4.1) ~{\rm MeV} \nn
10^3 L_4^r(M_\rho) &=& - 0.05 \pm 0.22 \nn
{\rm corr}(F_0,L_4^r) &=& - 0.996 ~.
\label{eq:fit3}
\end{eqnarray} 

The errors of $F_0$, $L_4$ are due to both lattice and
theoretical uncertainties. First, there are 
statistical errors of the lattice values for $F_\pi$ and
the meson masses and, in addition, the uncertainties of the inverse
lattice spacings. The dominant errors are those of the lattice
spacings and of $F_\pi$, whereas the errors of the
lattice masses are practically negligible. We have neglected unknown
correlations among the lattice data, thereby probably overestimating
the combined errors.  

In addition, we added the theoretical 
uncertainties related to $M$, $L_5$ and the $C_a$ in quadrature. 
Lattice and theoretical errors are of similar size. For
instance, keeping only the lattice errors, the error of $F_0$
moves from $4.1$ down to $2.8$ MeV. The $\chi^2/$dof is 0.5
(statistical errors only), suggesting once more that we have at least
not underestimated the errors. 

The two ellipses are roughly compatible with each other. The green
ellipse is lower because from the RBC/UKQCD data alone the fitted value of
$F_\pi$ is smaller than the experimental value. The value
for $L_4$ is consistent with large $N_c$ and with available lattice
results \cite{Aoki:2013ldr}. The result for
$F_0$ is more precise than existing phenomenological and lattice
determinations. It is somewhat bigger than expected
\cite{DescotesGenon:1999uh}, roughly of the same size as the
$SU(2)$ LEC $F$ in Eq.~(\ref{eq:F}). 

$F_0$ and $L_4$ in Eq.~(\ref{eq:fit3}) are compatible with the 
comparison between $SU(2)$ and $SU(3)$ to $O(p^6)$ \cite{Gasser:2007sg},
as indicated by the orange bands in Fig.~\ref{fig:F0FL4_34}. $C_{16}$ is
the only NNLO LEC  appearing in the relation between $F_0$ and $F$. As
always in this paper, we have used 
fit 10 \cite{Amoros:2001cp} for the NLO LECs. However,
unlike for our fit results (\ref{eq:fit3}), the orange bands in
Fig.~\ref{fig:F0FL4_34} are rather sensitive to the precise values of
the $L^r_i$. Therefore, although the consistency between the ellipses
and the lower orange band is manifest, it can hardly be used as
a determination of $C_{16}$.

Raising the range in pion masses to $M_\pi < 425$ MeV, two more
lattice points \cite{Aoki:2010dy} can be added. Repeating the fit with
the bigger sample moves the ellipses down, because with the original
data set of RBC/UKQCD the fitted value of $F_\pi$ comes out too low
\cite{Aoki:2010dy}. 

The strong anti-correlation between $F_0$ and $L_4$ persists because
the kaon masses in the RBC/UKQCD data are all close to the physical
kaon mass. Simulations with smaller kaon masses would not only be
welcome from the point of view of convergence of the chiral series
\cite{Bazavov:2009bb}, 
but they could also provide a better lever arm for reducing the
anti-correlation and the fit errors of $F_0$ and $L_4$. This
expectation is supported by the fact that
the quantity $F(M_\rho)$ defined in Eq.~(\ref{eq:Fmu}) can be
determined much better than $F_0$.

\section{Remarks on  $\mathbf{f^{K\pi}_+(0)}$} 
\label{sec:kl3}
\renewcommand{\theequation}{\arabic{section}.\arabic{equation}}
\setcounter{equation}{0}
The kaon semileptonic vector form factor at $t=0$ is a
crucial quantity for a precision determination of the CKM matrix element
$V_{us}$. Both approximations discussed here do not appear very
promising in this case.

First of all, unlike for $F_\pi$ and $F_K/F_\pi$, the chiral expansion
of $f^{K\pi}_+(0)$ shows a rather atypical behaviour. Due to the
Ademollo-Gatto theorem \cite{Ademollo:1964sr}, 
the $O(p^4)$ contribution of $- 0.0227$
\cite{Gasser:1984ux} is very small. On the basis of recent lattice
studies, which find $f^{K\pi}_+(0)=0.967$ with errors of less
than $1 \%$ \cite{Bazavov:2012cd,Boyle:2013gsa},
all higher-order contributions in CHPT would have to sum
up to about $- 1 \%$. On the other hand, the genuine two-loop contributions
at the usual scale $\mu=770$ MeV are positive and slightly bigger than
$1 \%$ \cite{Post:2001si,Bijnens:2003uy,Bernard:2009ds}, suggesting
that the remainder is about $- 2 \%$ to match the lattice value. In
other words, the remainder would have to be as big as the NLO
contribution, certainly not the typical behaviour for a chiral
expansion. 

In principle, Approximation I fulfills our criterion of
Sec.~\ref{sec:anapp} in differing from the full two-loop result
\cite{Post:2001si,Bijnens:2003uy,Bernard:2009ds} by less than $2
\%$. However, especially in view of the accuracy of recent lattice
studies claiming a precision of better than $1 \%$ for
$f^{K\pi}_+(0)$, the accuracy of Approximation I is simply not
sufficient in this case. Approximation II does not improve the situation.

To sum up, lattice determinations of $f^{K\pi}_+(0)$ seem to be able
to do without CHPT. Moreover, only the full NNLO expression may allow
for a meaningful extraction of LECs if at all \cite{Bernard:2009ds}.

\section{Conclusions} 
\label{sec:conc}
\renewcommand{\theequation}{\arabic{section}.\arabic{equation}}
\setcounter{equation}{0}
We summarize the main results of our work.

\begin{enumerate}
\item
Lattice QCD has become a major source of information for the
low-energy constants of CHPT. We have argued that the meson
decay constants $F_\pi$, $F_K$ are especially suited for
extracting chiral $SU(3)$ LECs of different chiral orders. The ratio
$F_K/F_\pi$ allows for a precise and stable determination of the NLO
LEC $L_5$. In addition, it gives access to some NNLO LECs although the
accuracy is of course more limited in that case. Phenomenological
analyses have had difficulties in determining the 
LEC $F_0$, the meson decay constant in the chiral $SU(3)$ limit. We
have shown that lattice data for $F_\pi$  allow for the
extraction of $F_0$ together with the NLO LEC $L_4$. The strong
anti-correlation between $F_0$ and $L_4$ observed in phenomenological
analyses can in principle be lifted by varying the lattice
masses. From a fit to the RBC/UKQCD data for $F_\pi$, we have
obtained a value for $F_0$ that is more precise than other presently
available determinations.
\item
Confronting present-day lattice data with chiral $SU(3)$ requires
chiral amplitudes to NNLO in most cases. Chiral $SU(3)$ amplitudes are
often rather unwieldy and mostly available in numerical form only. We
have therefore proposed large-$N_c$ motivated approximate NNLO
amplitudes that contain only one-loop functions. Unlike simpler
approximations as the double-log approximation, our amplitudes are
independent of the renormalization scale and can therefore be used to
extract LECs with the correct scale dependence. However,
approximations of NNLO amplitudes can only be
successful if the differences to the full amplitudes are at most of
the order of N$^3$LO contributions. We have checked that this
criterion can be fulfilled with our approximate amplitudes both for
$F_\pi$ and $F_K/F_\pi$. Therefore, we expect our results for the
different LECs to be as reliable as CHPT to NNLO, $O(p^6)$, 
permits. Although our general criterion is also satisfied for the kaon 
semileptonic form factor at $t=0$, the approximate expression for
$f^{K\pi}_+(0)$ is not precise enough compared to recent lattice data. 
\end{enumerate}     

The main purpose of this work has been to encourage lattice groups to
use NNLO amplitudes in chiral $SU(3)$ that are more user friendly
than the full expressions and yet are reliable enough to provide more
insight than NLO amplitudes with polynomial corrections.

%\vspace*{.8cm} 

\paragraph{Acknowledgements}
We are grateful to V\'eronique Bernard, Claude Bernard, Gilberto
Colangelo, Laurent Lellouch,  Heiri Leutwyler,
Emilie Passemar and Lothar Tiator for helpful comments and
suggestions. We are indebted to
Elvira G\'amiz for helping us to understand lattice data. 
Special thanks are due to Hans Bijnens for suggesting several
substantial improvements of
the original manuscript and for making the full results of
Ref.~\cite{Amoros:1999dp} accessible to us.   
P.M. acknowledges support from the Deutsche Forschungsgemeinschaft DFG
through the Collaborative Research Center ``The Low-Energy Frontier of
the Standard Model'' (SFB 1044).

\vspace*{1cm}
%\newpage

\appendix
\renewcommand{\theequation}{\Alph{section}.\arabic{equation}}
\renewcommand{\thetable}{\Alph{section}.\arabic{table}}
\setcounter{equation}{0}
\setcounter{table}{0}

\section{Generating functional of $\mathbf{O(p^6)}$}
\label{app:gfp6}
\noindent
In this appendix we rederive the generating functional of $O(p^6)$ in
the form used in Ref.~\cite{Ecker:2010nc}. It is a more explicit
version of the derivation in Ref.~\cite{Bijnens:1999hw}. 
 
The generating functional $Z_6$ is shown pictorially in
Fig.~\ref{fig:p6diag}. The various contributions to
$Z_6$ are always understood as functionals of the classical
field, the solution of the lowest-order field equations.

As discussed in Ref.~\cite{Bijnens:1999hw}, the sum of the
reducible diagrams c, e, f leads to a finite and scale independent
functional with the conventional choice of chiral Lagrangians. The
contributions from diagrams a, b and d are divergent. The sum
$Z_6^{\rm a+b+d}$ is
still divergent, but the divergence takes the form of a local
functional that is canceled by the divergent part of the tree-level
functional $Z_6^{\rm g}$ in terms of the LECs $C_a$ of $O(p^6)$.  

We first consider the irreducible two-loop diagrams a, b. In d
dimensions, the corresponding functional has the form
\begin{eqnarray} 
Z_6^{\rm a+b} &=& \int \!\! d^dx (c\mu)^{2(d-4)} \left\{ \Lambda^2
  \sum_a \alpha_a O_a(x) + \displaystyle\frac{\Lambda}{(4 \pi)^2}
  \left[\sum_a \left(\beta_a + \alpha_a \ln{M^2/\mu^2}\right) O_a(x) + D(x;M)
  \right] \right. \no \\[.1cm]  
&& \hspace*{-1.3cm} + \left. \displaystyle\frac{1}{(4 \pi)^4} \left[ E(x;M) +
  \ln^2{M^2/\mu^2} \sum_a \displaystyle\frac{\alpha_a}{2} O_a(x)
  + \ln{M^2/\mu^2} \left(\sum_a \beta_a O_a(x) + D(x;M) \right) \right] 
  \right\} 
\label{eq:a+b}
\end{eqnarray} 
with the divergence factor
\begin{equation} 
\Lambda=\dfrac{1}{(4 \pi)^2 (d-4)} ~. 
\label{Lambda}
\end{equation} 
The monomials $O_a(x) ~(a=1,\dots,94)$ define the chiral $SU(3)$
Lagrangian of $O(p^6)$ \cite{Bijnens:1999sh}. $D(x;M)$ and $E(x;M)$
are nonlocal functionals. The mass $M$ is introduced to make the
dependence on the renormalization scale $\mu$ explicit. At this stage,
the functional $Z_6^{\rm a+b}$ is independent of $M$. The scheme
dependent constant $c$  is conventionally chosen as in Eq.~(2.22) of
Ref.~\cite{Bijnens:1999hw}. Eq.~(\ref{eq:a+b}) is equivalent to Eq.~(2.39) in
Ref.~\cite{Bijnens:1999hw} but the renormalization group equations 
(2.40) \cite{Bijnens:1999hw} have already been taken into account. In
other words, the scale independence of $Z_6^{\rm a+b}$ is made
explicit implying 
\begin{eqnarray} 
\mu \displaystyle\frac{\partial \alpha_a}{\partial \mu}=0 ~, \qquad &
\qquad  
\mu \displaystyle\frac{\partial \beta_a}{\partial \mu} = 0 \no \\[.1cm] 
\mu \displaystyle\frac{\partial D(x;M)}{\partial \mu} = 0 ~, \qquad &
\qquad 
\mu \displaystyle\frac{\partial E(x;M)}{\partial \mu} = 0 ~.
\end{eqnarray}  

The general structure of the irreducible one-loop functional d is 
\begin{equation} 
Z_6^{\rm d} =  \int \!\! d^dx (c\mu)^{(d-4)} \sum_{i=1}^{10} L_i(d) Y_i(x,d)
\end{equation} 
where the LECs of $O(p^4)$ are decomposed as 
\begin{equation}
L_i(d) = (c\mu)^{d-4} \left[\Gamma_i \Lambda + L_i^r(\mu,d) \right]~.
\end{equation}
Adopting the renormalization conventions of
Ref.~\cite{Bijnens:1999hw}, the LECs $L_i^r(\mu,d)$ are 
not expanded around $d=4$. Scale independence of the $L_i(d)$ then
implies 
\begin{equation}
\mu \displaystyle\frac{\partial L_i^r(\mu,d)}{\partial \mu} = -
\displaystyle\frac{\Gamma_i}{(4 \pi)^2} - (d-4) L_i^r(\mu,d) ~. 
\end{equation}
Because of the divergence in $L_i(d)$ one must keep track of 
terms of $O(d-4)$ in $Y_i(x,d)$. With hindsight, these functionals can
be written as
\begin{eqnarray} 
Y_i(x,d) &=& \left(\Lambda + \displaystyle\frac{1}{2 (4\pi)^2}
\ln{M^2/\mu^2}  \right) \sum_a \eta_a^i O_a(x) \\[.1cm] 
&& \hspace*{-1.5cm} + \displaystyle\frac{1}{(4 \pi)^2} \left\{H_i(x,d;M) + (d-4)
  \left[\displaystyle\frac{1}{8} \ln^2{M^2/\mu^2} \sum_a \eta_a^i
    O_a(x) + \displaystyle\frac{1}{2} \ln{M^2/\mu^2} ~H_i(x,d;M)  
\right] \right\} ~. \no
\end{eqnarray} 
The scale independence of $(c\mu)^{d-4} \,Y_i(x,d)$ implies that the
coefficients $\eta_a^i$ are scale independent and that the functionals 
$H_i(x,d;M)$ satisfy the renormalization group equations 
\begin{equation} 
\mu \displaystyle\frac{\partial H_i(x,d)}{\partial \mu} =
O[(d-4)^2]~. 
\label{eq:RGE_H}
\end{equation}
Putting everything together, we obtain (using for convenience from now
on the summation convention for both indices a, i)
\begin{eqnarray} 
Z_6^{\rm d} &=&  \int \!\! d^dx (c\mu)^{2(d-4)} \left\{ \Lambda^2 \,
\Gamma_i \eta_a^i O_a(x)  \right.  \\[.1cm]
&& + \displaystyle\frac{\Lambda}{(4 \pi)^2} \left[(4 \pi)^2 L_i^r(\mu,d) 
\, \eta_a^i O_a(x) +
  \displaystyle\frac{1}{2} \ln{M^2/\mu^2}\, \Gamma_i \eta_a^i
  O_a(x) +  \Gamma_i H_i(x,d;M) \right]  \no \\[.2cm]     
&& + \left. \displaystyle\frac{1}{(4 \pi)^4}
\left[\displaystyle\frac{1}{8} \ln^ 2{M^2/\mu^2}\, \Gamma_i \eta_a^i
  O_a(x) +   \displaystyle\frac{1}{2} \ln{M^2/\mu^2}\, \Gamma_i
  H_i(x,4;M) \right. \right. \no \\[.1cm]  
&& + \left. \left. \displaystyle\frac{1}{2} \ln{M^2/\mu^2}\,(4 \pi)^2
  L_i^r(\mu,4) \eta_a^i O_a(x) +  (4 \pi)^2 L_i^r(\mu,4) H_i(x,4;M)
  \right] + O(d-4) \right\} ~. \no
\end{eqnarray}  
Altogether, the irreducible contributions sum up to the functional
\begin{eqnarray} 
Z_6^{\rm a+b+d} &=&  \int \!\! d^dx (c\mu)^{2(d-4)} \left\{ \Lambda^2
\left[\alpha_a + \Gamma_i \eta_a^i \right] O_a(x)
\right.  \label{eq:irred} \\[.2cm] 
&& + \displaystyle\frac{\Lambda}{(4 \pi)^2} \left[\beta_a O_a(x) +
  \ln{M^2/\mu^2} \left(\alpha_a + \displaystyle\frac{1}{2} \Gamma_i
  \eta_a^i  \right) O_a(x) \right. \no \\[.1cm] 
&& + \left. D(x;M) +  \Gamma_i H_i(x,d;M) + (4 \pi)^2 L_i^r(\mu,d) \, 
  \eta_a^i O_a(x) \right] \no \\[.2cm]     
&& + \displaystyle\frac{1}{(4 \pi)^4}
\left[E(x;M) +
  \ln^2{M^2/\mu^2}\, \displaystyle\frac{\alpha_a}{2} O_a(x)
  + \ln{M^2/\mu^2} \left( \beta_a O_a(x) + D(x;M) \right)
  \right. \no \\[.1cm] 
&& + 
\displaystyle\frac{1}{8} \ln^ 2{M^2/\mu^2}\, \Gamma_i \eta_a^i
  O_a(x) +   \displaystyle\frac{1}{2} \ln{M^2/\mu^2}\, \Gamma_i
  H_i(x,4;M)  \no \\[.1cm]  
&& + \left. \left. \displaystyle\frac{1}{2} \ln{M^2/\mu^2}\,(4 \pi)^2
  L_i^r(\mu,4) \eta_a^i O_a(x) +  (4 \pi)^2 L_i^r(\mu,4) H_i(x,4;M)
  \right] + O(d-4) \right\} \,. \no
\end{eqnarray}  
The double-pole divergence functional is automatically local. In order
to cancel the divergences with the local functional $Z_6^{\rm g}$, also the
single-pole divergences in (\ref{eq:irred}) must be local. Absence of
the logarithmic terms implies the 94 Weinberg conditions \cite{Weinberg:1978kz}
\begin{equation} 
\alpha_a = - \displaystyle\frac{1}{2} \Gamma_i \eta_a^i ~. 
\label{eq:weinberg}
\end{equation}
Moreover, the non-local functional $D(x;M)$ must be canceled by $\Gamma_i
  H_i(x,4;M)$. More precisely, renormalization theory requires that
\begin{equation} 
D(x;M) + \Gamma_i H_i(x,4;M) = \Delta \beta_a \,O_a(x)~,
\label{eq:nonlocal}
\end{equation}
i.e., that the sum of the two terms is local. In
Ref.~\cite{Bijnens:1999hw} it was found
that the cancellation is complete: $\Delta \beta_a = 0$.  

Using Eqs.~(\ref{eq:weinberg}) and (\ref{eq:nonlocal}) (with $\Delta
\beta_a = 0$), the irreducible functional is given by
\begin{eqnarray} 
Z_6^{\rm a+b+d} &=&  \int \!\! d^dx (c\mu)^{2(d-4)} \left\{ 
\displaystyle\frac{\Lambda^2}{2}  \Gamma_i \eta_a^i O_a(x)  
 + \displaystyle\frac{\Lambda}{(4 \pi)^2} \left[\beta_a O_a(x) +
   (4 \pi)^2 L_i^r(\mu,d) \, \eta_a^i O_a(x) \right] \right. \no \\[.2cm]     
&& + \displaystyle\frac{1}{(4 \pi)^4}
\left[- \displaystyle\frac{1}{8} \ln^ 2{M^2/\mu^2}\, \Gamma_i \eta_a^i
  O_a(x) + \displaystyle\frac{1}{2} \ln{M^2/\mu^2}\,(4 \pi)^2
  L_i^r(\mu,4) \eta_a^i O_a(x) \right. \no \\[.1cm] 
&& + \ln{M^2/\mu^2}\, \beta_a O_a(x) +
H_i(x,4;M) \left((4 \pi)^2 L_i^r(\mu,4) - \displaystyle\frac{\Gamma_i }{2}
\ln{M^2/\mu^2}  \right)  \no \\[.1cm] 
&& + \left. \left. E(x;M) + \Gamma_i H_i^\prime(x,4;M)\right] + O(d-4)
\right\} ~.
\label{eq:irredfinal} 
\end{eqnarray} 
The functional $H_i^\prime(x,4;M)$ is defined by
the Taylor expansion
\begin{equation}
H_i(x,d;M) = H_i(x,4;M) + (d-4) H_i^\prime(x,4;M) + O[(d-4)^2]
\end{equation}  
and it is scale independent because of Eq.~(\ref{eq:RGE_H}).

Now we can render the complete functional finite by adding the tree-level
functional of $O(p^6)$ (in the notation of Eq.~(4.9) in
Ref.~\cite{Bijnens:1999hw}): 
\begin{eqnarray}
Z_6^{\rm g} &=&  \int \!\! d^dx \,C_a(d) O_a(x) \label{eq:treep6}
\\[.1cm] 
&=&  \int \!\! d^dx (c\mu)^{2(d-4)}\left[ C_a^r(\mu,d) -
  \displaystyle\frac{\Gamma_a^{(2)}}{F^2} \Lambda^2 -
  \displaystyle\frac{1}{F^2}  
\left(\Gamma_a^{(1)} + \Gamma_a^{(L)}(\mu,d) \right) \Lambda \right]
 O_a(x) ~.\no
\end{eqnarray}
Comparing Eqs.~(\ref{eq:irredfinal}) and (\ref{eq:treep6}),
the divergences are canceled with
\begin{eqnarray}
\Gamma_a^{(2)} = \displaystyle\frac{F^2}{2} \Gamma_i \eta_a^i ~,
\qquad &  \Gamma_a^{(1)} = \displaystyle\frac{F^2}{(4\pi)^2} \beta_a~,
\qquad &  \Gamma_a^{(L)}(\mu,d) = F^2 \,L_i^r(\mu,d) \, \eta_a^i ~.
\end{eqnarray}
The coefficients
$\Gamma_a^{(1)},\,\Gamma_a^{(2)},\,\Gamma_a^{(L)}(\mu,d)$ are listed
in  Table II of App.~C in Ref.~\cite{Bijnens:1999hw}.

Summing up the diagrams a, b, d and g, the limit $d \to 4$ can now be
taken to arrive at the final result
\begin{eqnarray}
Z_6^{\rm a+b+d+g} &=&  \int \!\! d^4x \left\{ C_a^r(\mu) O_a(x) +
\displaystyle\frac{1}{4 F^2} \left(4\, \Gamma_a^{(1)} \,L(\mu)  -
\Gamma_a^{(2)} \,L^2(\mu) + 2\, \Gamma_a^{(L)}(\mu) L(\mu) \right) O_a(x)
\right. \no \\[.2cm]
&& + \left. \displaystyle\frac{1}{(4\pi)^2} \left[L_i^r(\mu) - 
  \displaystyle\frac{\Gamma_i }{2} L(\mu) \right] H_i(x;M) 
+ \displaystyle\frac{1}{(4 \pi)^4} K(x;M) \right\}
\label{eq:Zfinite}
\end{eqnarray}
with the chiral log $L(\mu)$ defined in Eq.~(\ref{eq:clog}) and with
\begin{eqnarray}
C_a^r(\mu) = C_a^r(\mu,4)~, \quad  L_i^r(\mu) = L_i^r(\mu,4)~, \quad 
\Gamma_a^{(L)}(\mu) = \Gamma_a^{(L)}(\mu,4) \no \\[.2cm] 
 \quad H_i(x;M)=H_i(x,4;M), \quad  K(x;M) = E(x;M) + \Gamma_i
 H_i^\prime(x,4;M)~. 
\end{eqnarray}
The scale dependence is contained in
$C_a^r(\mu),\,L_i^r(\mu),\,L(\mu)$. The functionals 
$O_a(x)$, $H_i(x;M)$ and $K(x;M)$ are independent of $\mu$. 
Scale independence of the complete functional (\ref{eq:Zfinite}) can
be checked with the help of the renormalization group equations (4.5) in
Ref.~\cite{Bijnens:1999hw}: 
\begin{equation}
\mu\displaystyle\frac{d C_a^r(\mu)}{d\mu}=
\displaystyle\frac{1}{(4\pi)^2 F^2}\left[2 \Gamma^{(1)}_a +
  \Gamma^{(L)}_a(\mu) \right] ~.
\end{equation}
As already mentioned, the sum of reducible diagrams c, e, f is finite
and scale independent by itself. It can be written in the form  
\begin{eqnarray}
Z_6^{\rm c+e+f} &=&  \int \!\! d^4x\,d^4y \left[\left(L_i^r(\mu) - 
  \displaystyle\frac{\Gamma_i }{2} L(\mu) \right)
  P_{i,\alpha}(x) + F_\alpha(x;M) \right]G_{\alpha,\beta}(x,y) \no
  \\[.1cm]  
&& \times \left[\left(L_j^r(\mu) - 
  \displaystyle\frac{\Gamma_j }{2} L(\mu) \right) P_{j,\beta}(y) +
  F_\beta(y;M) \right] ~. 
\label{eq:reduc}
\end{eqnarray}
The derivatives of the monomials $P_i(x)$ defining the chiral
Lagrangian of $O(p^4)$ with respect to the fields
$\vp_\alpha$ ($\alpha=1, \dots,8$) are denoted $P_{i,\alpha}(x)$. The
$F_\alpha(x;M)$ are finite and scale independent one-loop functionals. The
propagator $G_{\alpha,\beta}(x,y)$ is again a functional of the
classical field. Although the functional (\ref{eq:reduc}) is nonlocal
in general, it contributes in many cases of interest to wave function,
mass and decay constant renormalization only. 

The complete generating functional of $O(p^6)$ is then given by the
sum
\begin{equation}
Z_6 = Z_6^{\rm a+b+d+g} + Z_6^{\rm c+e+f}~.
\label{eq:Ztotal}
\end{equation}
Once again, it is independent of both scales $\mu$ and $M$.

\newpage

\setcounter{equation}{0}
\setcounter{table}{0}

\section{Approximation II for $\mathbf{F_K/F_\pi}$}
\label{app:fkfpi}

The original Approximation I for  $F_K/F_\pi$ was given in the
appendix of Ref.~\cite{Ecker:2010nc}. In Approximation II discussed in
Sec.~\ref{sec:fkfpi}, there is an additional contribution of $O(p^6)$
denoted $R_6^{\rm ext}$ below. The complete result for $F_K/F_\pi$ is 
\begin{eqnarray} 
F_K/F_\pi &=& 1 + R_4 + R_6 + R_6^{\rm ext} ~,
\label{eq:fkfpi}
\\[.3cm] 
      F_0^2 \,R_4 &=&  4 \,(\as{\circ}{M}_K^2 - \as{\circ}{M}_\pi^2) \,L_5 
        - 5 \,\overline{A}(\as{\circ}{M}_\pi,\mu)/8
       + \overline{A}(\as{\circ}{M}_K,\mu)/4 
       +  3 \,\overline{A}(\as{\circ}{M}_\eta,\mu)/8 ~,  \\[.3cm]
      F_0^4 \,R_6 &=& 8\,F_0^2 (\as{\circ}{M}_K^2 - \as{\circ}{M}_\pi^2) \left( 2 \,\as{\circ}{M}_K^2\,
     (C_{14} + C_{15}) + \as{\circ}{M}_\pi^2 \,(C_{15} + 2 \,C_{17}) \right) \no
      \\[.1cm] 
&+& (\as{\circ}{M}_K^2 - \as{\circ}{M}_\pi^2)\, \left( - 32 \,(\as{\circ}{M}_\pi^2 + 2 \,\as{\circ}{M}_K^2) L_4\,L_5  
- 8 \,(3 \,\as{\circ}{M}_\pi^2 + \as{\circ}{M}_K^2) L_5^2  \right. \no \\[.1cm]  
&& + \left. (25 \,\as{\circ}{M}_\pi^2 + 17 \,\as{\circ}{M}_K^2)L^2 /32
      \right) \no \\[.1cm]
&+& \displaystyle\frac{(\as{\circ}{M}_K^2 - \as{\circ}{M}_\pi^2)}{(4\pi)^2} \left( -
      2 \,(\as{\circ}{M}_\pi^2 + \as{\circ}{M}_K^2) L_1 -  (\as{\circ}{M}_\pi^2 + \as{\circ}{M}_K^2) L_2
 - (5 \,\as{\circ}{M}_\pi^2 + \as{\circ}{M}_K^2) L_3/18  \right. \no \\[.1cm] 
&&+\left. 6 \,(\as{\circ}{M}_\pi^2 + 2 \,\as{\circ}{M}_K^2) L_4 +  
(14 \,\as{\circ}{M}_\pi^2 + 22 \,\as{\circ}{M}_K^2) L_5/3   - 12\,(\as{\circ}{M}_\pi^2 + 2 \,\as{\circ}{M}_K^2) L_6  
\right. \no \\[.1cm]
&& + \left. 16 \,(\as{\circ}{M}_\pi^2 - \as{\circ}{M}_K^2) L_7 -  4\,(\as{\circ}{M}_\pi^2 + 5 \,\as{\circ}{M}_K^2) L_8 
+ (313 \,\as{\circ}{M}_\pi^2 + 271 \,\as{\circ}{M}_K^2) L/288
\right)  \no \\[.1cm]
&+& 5 \,\overline{A}(\as{\circ}{M}_\pi,\mu)^2/8  - \overline{A}(\as{\circ}{M}_K,\mu)^2/8
   + \overline{A}(\as{\circ}{M}_\pi,\mu) \,\overline{A}(\as{\circ}{M}_K,\mu)/16  \no \\[.1cm]    
&-& 3 \,\overline{A}(\as{\circ}{M}_\pi,\mu) \,\overline{A}(\as{\circ}{M}_\eta,\mu)/8 
 - 3 \,\overline{A}(\as{\circ}{M}_K,\mu) \,\overline{A}(\as{\circ}{M}_\eta,\mu)/16  \no \\[.1cm]
&+& \overline{A}(\as{\circ}{M}_\pi,\mu) \left( 4 \,\as{\circ}{M}_\pi^2 L_1 + 10 \,\as{\circ}{M}_\pi^2 L_2 +
 13 \,\as{\circ}{M}_\pi^2 L_3/2 + 10 \,(\as{\circ}{M}_\pi^2 + 2 \,\as{\circ}{M}_K^2) L_4  \right. \no
      \\[.1cm] 
&& + \left. (19 \,\as{\circ}{M}_\pi^2 - 5 \,\as{\circ}{M}_K^2) L_5/2 -  10 \,(\as{\circ}{M}_\pi^2 + 2
      \,\as{\circ}{M}_K^2) L_6 -  10 \,\as{\circ}{M}_\pi^2 L_8 \right. \no \\[.1cm]
&&- \left. (361 \,\as{\circ}{M}_\pi^2 + 131 \,\as{\circ}{M}_K^2) L/288 \right)\no \\[.1cm]
&+& \overline{A}(\as{\circ}{M}_K,\mu) \left(  - 4 \,\as{\circ}{M}_K^2 L_1 - 10 \,\as{\circ}{M}_K^2 L_2  
- 5 \,\as{\circ}{M}_K^2 L_3 - 4 \,(\as{\circ}{M}_\pi^2 + 2 \,\as{\circ}{M}_K^2) L_4 
\right. \no \\[.1cm]
&&- \left. (\as{\circ}{M}_\pi^2 + \as{\circ}{M}_K^2) L_5 +  4\,(\as{\circ}{M}_\pi^2 + 2 \,\as{\circ}{M}_K^2) L_6 + 4 \,\as{\circ}{M}_K^2 L_8 
+ (59 \,\as{\circ}{M}_\pi^2  + 115 \,\as{\circ}{M}_K^2) L/144 \right) \no \\[.1cm]
&+& \overline{A}(\as{\circ}{M}_\eta,\mu) (\as{\circ}{M}_K^2 - \as{\circ}{M}_\pi^2)/\as{\circ}{M}_\eta^2 \left(- 9 \,\as{\circ}{M}_\pi^2 L_7 
- 3 \,\as{\circ}{M}_\pi^2 L_8 + 5 \,\as{\circ}{M}_\pi^2 L/32 \right) \no \\[.1cm]
&+& \overline{A}(\as{\circ}{M}_\eta,\mu) \left( (\as{\circ}{M}_\pi^2/2 - 2 \,\as{\circ}{M}_K^2) L_3 
- 6\,(\as{\circ}{M}_\pi^2 + 2 \,\as{\circ}{M}_K^2) L_4 - (7 \,\as{\circ}{M}_\pi^2 + 23 \,\as{\circ}{M}_K^2) L_5/6 \right.
\no \\[.1cm] 
&&+ \left. 6 \,(\as{\circ}{M}_\pi^2 + 2 \,\as{\circ}{M}_K^2) L_6 + 3 \,(3 \,\as{\circ}{M}_\pi^2 \as{\circ}{M}_K^2/\as{\circ}{M}_\eta^2 
- 7 \,\as{\circ}{M}_\pi^2 + 4 \,\as{\circ}{M}_K^2) L_7 \right.\no \\[.1cm] 
&&+ \left. 3 \,(\as{\circ}{M}_\pi^2 \as{\circ}{M}_K^2/\as{\circ}{M}_\eta^2 - 3 \,\as{\circ}{M}_\pi^2 + 4 \,\as{\circ}{M}_K^2) L_8
      \right. \no \\[.1cm] 
&& \left. -  (15 \,\as{\circ}{M}_\pi^2 \as{\circ}{M}_K^2/\as{\circ}{M}_\eta^2 - 44 \,\as{\circ}{M}_\pi^2 -
19 \,\as{\circ}{M}_K^2) L /96  \right) ~,
\end{eqnarray} 
\newpage
\begin{eqnarray}   
F_0^4\,R_6^{\rm ext} &=&
- \displaystyle\frac{(\as{\circ}{M}_K^2-\as{\circ}{M}_\pi^2)}{(4\pi)^2}\,(17 \as{\circ}{M}_K^2+ 10
\as{\circ}{M}_\pi^2)\,L/48  
 - (\as{\circ}{M}_K^2-\as{\circ}{M}_\pi^2)\,(11 \as{\circ}{M}_K^2+7 \as{\circ}{M}_\pi^2)\,L^2/96 \no \\[.1cm] 
&+& \displaystyle\frac{(\as{\circ}{M}_K^2-\as{\circ}{M}_\pi^2)}{(4\pi)^2}\,(10 \as{\circ}{M}_K^2 +
17 \as{\circ}{M}_\pi^2)/4608 
 +\displaystyle\frac{(\as{\circ}{M}_K^2-\as{\circ}{M}_\pi^2)}{(4\pi)^4}\,(17 \as{\circ}{M}_K^2 + 
  19 \as{\circ}{M}_\pi^2)/384 \no \\[.1cm] 
&+&  \overline{A}(\as{\circ}{M}_\pi,\mu)\left\{- (35 \as{\circ}{M}_\pi^2 + 49 \as{\circ}{M}_K^2)\,L/288+
\displaystyle\frac{1}{(4\pi)^2}\,(8 \as{\circ}{M}_\pi^2 + \as{\circ}{M}_K^2)/32\right\} \no
      \\[.1cm] 
&-& 
\overline{A}(\as{\circ}{M}_\pi,\mu)^2\,(19 + 20 \as{\circ}{M}_K^2/\as{\circ}{M}_\pi^2)/128 +
3 \overline{A}(\as{\circ}{M}_\pi,\mu) \overline{A}(\as{\circ}{M}_K,\mu)/32 \no \\[.1cm] 
&+&  \overline{A}(\as{\circ}{M}_\pi,\mu) \overline{A}(\as{\circ}{M}_\eta,\mu)\,(7
+36 \as{\circ}{M}_\pi^2/\as{\circ}{M}_\eta^2)/192 \no \\[.1cm] 
&+& \overline{A}(\as{\circ}{M}_K,\mu)\left\{(121 \as{\circ}{M}_\pi^2-115 \as{\circ}{M}_K^2)\,L/144 +
\displaystyle\frac{1}{(4\pi)^2}\,(- 4 \as{\circ}{M}_\pi^2+7 \as{\circ}{M}_K^2)/16 \right\}
      \no \\[.2cm] 
&+& \overline{A}(\as{\circ}{M}_K,\mu)^2\,(3 + 10 \as{\circ}{M}_\pi^2/ \as{\circ}{M}_K^2)/32 -
\overline{A}(\as{\circ}{M}_K,\mu) \overline{A}(\as{\circ}{M}_\eta,\mu)\,(71 + 12
\as{\circ}{M}_\pi^2/\as{\circ}{M}_\eta^2)/96 \no \\[.1cm] 
&+& \overline{A}(\as{\circ}{M}_\eta,\mu)\left\{(- 12 \as{\circ}{M}_\pi^2 \as{\circ}{M}_K^2/\as{\circ}{M}_\eta^2
- 23 \as{\circ}{M}_\pi^2+ 18 \as{\circ}{M}_\pi^4/\as{\circ}{M}_\eta^2 + 41 \as{\circ}{M}_K^2)\,L/96
\right. \no \\[.1cm] 
&& \hspace*{.3cm} + \left.\displaystyle\frac{1}{(4\pi)^2}\,(4 \as{\circ}{M}_\pi^2-19
\as{\circ}{M}_K^2)/32\right\} 
+ \overline{A}(\as{\circ}{M}_\eta,\mu)^2\,(56 + \as{\circ}{M}_\pi^2/\as{\circ}{M}_\eta^2)/128~. 
\end{eqnarray} 
%%%%%%%%%%%%%%%%%%%%%%%%%%%%%%%%%%%%%%%%%%%%%%%%%%%%%%%%%%%%%%%%%%%
%%     Approximation II                                         %%%
%%%%%%%%%%%%%%%%%%%%%%%%%%%%%%%%%%%%%%%%%%%%%%%%%%%%%%%%%%%%%%%%%%%
We use $L_i=L_i^r(\mu)$, $C_a=C_a^r(\mu)$, $L=L(\mu)$ 
for a compact representation. The masses $\as{\circ}{M}_\alpha$
are the lowest-order meson masses of $O(p^2)$. Since we work to
$O(p^6)$, substituting the lowest-order masses $\as{\circ}{M}_\alpha$
by the actual lattice masses generates an additional contribution of
$O(p^6)$ \cite{Gasser:1984gg}.  
$F_0$ is the meson decay constant in the chiral
$SU(3)$ limit and the chiral log $L$ is defined in Eq.~(\ref{eq:clog}). The
loop function $\overline{A}(M_\alpha,\mu)$ is defined as
\begin{equation}
\overline{A}(M_\alpha,\mu)
=\displaystyle\frac{M_\alpha^2}{(4\pi)^2}
\ln{\displaystyle\frac{\mu^2}{M_\alpha^2}}~. 
\label{eq:1loop} 
\end{equation}

\setcounter{equation}{0}
\setcounter{table}{0}

\section{Approximation I for $\mathbf{F_\pi}$}
\label{app:fpif0}

In the approximation defined by the functional
(\ref{eq:approxI}), $F_\pi$ assumes the form
\begin{eqnarray}
F_\pi &=& F_0 \no \\[.3cm] 
&+& F_0^{-1}\left\{4\,(2\,\as{\circ}{M}_K^2+\as{\circ}{M}_\pi^2)\,L_4 + 4\,\as{\circ}{M}_\pi^2\,L_5 +
\overline{A}(\as{\circ}{M}_\pi,\mu)+ \overline{A}(\as{\circ}{M}_K,\mu)/2\right\} \no \\[.3cm] 
&+&
F_0^{-1}\left\{8\,\as{\circ}{M}_\pi^4(C_{14} + C_{15}+ 3 C_{16} + C_{17})
+ 16\,\as{\circ}{M}_\pi^2\,\as{\circ}{M}_K^2(C_{15} - 2 C_{16})
+ 32\,\as{\circ}{M}_K^4C_{16} \right\} \no \\[.1cm] 
&+&
\displaystyle\frac{F_0^{-3}}{(4\pi)^2}
  \,(284\,\as{\circ}{M}_\pi^2\,\as{\circ}{M}_K^2+
525\,\as{\circ}{M}_\pi^4+ 608\,\as{\circ}{M}_K^4)\,L/288 \no \\[.1cm] 
&+& F_0^{-3}\,(- 34\,\as{\circ}{M}_\pi^2\,\as{\circ}{M}_K^2+185\,\as{\circ}{M}_\pi^4 +
164\,\as{\circ}{M}_K^4)\,L^2/144 \no \\[.1cm] 
&+&
\displaystyle\frac{F_0^{-3}}{(4
  \pi)^2}\left\{-2\,\as{\circ}{M}_\pi^4\,L_1 + (8\,\as{\circ}{M}_\pi^2\,\as{\circ}{M}_K^2
 - 37\,\as{\circ}{M}_\pi^4 - 52\,\as{\circ}{M}_K^4)\,L_2/9\right. \no \\[.1cm]
 && + (8\,\as{\circ}{M}_\pi^2\,\as{\circ}{M}_K^2 -28 \,\as{\circ}{M}_\pi^4 - 43\,\as{\circ}{M}_K^4)\,L_3/27 
 + 4\,(5\,\as{\circ}{M}_\pi^2\,\as{\circ}{M}_K^2 + 2\,\as{\circ}{M}_\pi^4 + 2\,\as{\circ}{M}_K^4)\,L_4 \no \\[.1cm] 
 && + \left.4\,(2\,\as{\circ}{M}_\pi^4 + \as{\circ}{M}_K^4)\,L_5 -
8\,(5\,\as{\circ}{M}_\pi^2\,\as{\circ}{M}_K^2 + 2\,\as{\circ}{M}_\pi^4 + 2\,\as{\circ}{M}_K^4)\,L_6 
 - 8\,(2\,\as{\circ}{M}_\pi^4 + \,\as{\circ}{M}_K^4)\,L_8 \right\} \no \\[.2cm]
&+&  F_0^{-3}\left\{-8\,(4\,\as{\circ}{M}_\pi^2\,\as{\circ}{M}_K^2 + \as{\circ}{M}_\pi^4
 +4\,\as{\circ}{M}_K^4)\,L_4^2
-16\,(2\,\as{\circ}{M}_\pi^2\,\as{\circ}{M}_K^2 + \as{\circ}{M}_\pi^4)\,L_4 L_5 - 8\, \as{\circ}{M}_\pi^4 \,L_5^2\right\}
\no \\[.2cm] 
&+& F_0^{-3} \overline{A}(\as{\circ}{M}_\pi,\mu)\left\{- 28\,\as{\circ}{M}_\pi^2\,L_1 -
16\,\as{\circ}{M}_\pi^2\,L_2 - 14\,\as{\circ}{M}_\pi^2\,L_3 -
24\,\as{\circ}{M}_K^2\,L_4 \right. \no \\[.1cm] 
&& - \left. 6\,\as{\circ}{M}_\pi^2\,L_5 + 16\,(\as{\circ}{M}_\pi^2+ 2\,\as{\circ}{M}_K^2)\,L_6 +
16\,\as{\circ}{M}_\pi^2\,L_8 + \,(359\,\as{\circ}{M}_\pi^2+ 40\,\as{\circ}{M}_K^2)\,L/144\right\} \no
\\[.1cm] 
&+& F_0^{-3} \overline{A}(\as{\circ}{M}_K,\mu)\left\{-32\,\as{\circ}{M}_K^2\,L_1 - 8\,\as{\circ}{M}_K^2\,L_2
- 10\,\as{\circ}{M}_K^2\,L_3 + 2\,(-3\,\as{\circ}{M}_\pi^2+ 2\,\as{\circ}{M}_K^2)\,L_4 \right. \no \\[.1cm] 
&& + \left. 2\,(\as{\circ}{M}_\pi^2 - 2\,\as{\circ}{M}_K^2)\,L_5 + 8\,(\as{\circ}{M}_\pi^2+ 2\,\as{\circ}{M}_K^2)\,L_6
+ 8\,\as{\circ}{M}_K^2\,L_8 + \,(-11\,\as{\circ}{M}_\pi^2+ 62\,\as{\circ}{M}_K^2)\,L/36\right\}
\no \\[.1cm] 
&+& F_0^{-3} \overline{A}(\as{\circ}{M}_\eta,\mu)\left\{8\,(\as{\circ}{M}_\pi^2- 4\,\as{\circ}{M}_K^2)\,L_1/3 
+ 2\,(\as{\circ}{M}_\pi^2 - 4\,\as{\circ}{M}_K^2)\,L_2/3+
2\,(\as{\circ}{M}_\pi^2 - 4\,\as{\circ}{M}_K^2)\,L_3/3 \right. \no \\[.1cm] 
&& + \left. 4\,(- \as{\circ}{M}_\pi^2+ 4\,\as{\circ}{M}_K^2)\,L_4/3+
2\,\as{\circ}{M}_\pi^2\,L_5/3+ \,(- 11\,\as{\circ}{M}_\pi^2+ 20\,\as{\circ}{M}_K^2)\,L/48 \right\}~.
\label{eq:fpiI}
\end{eqnarray} 
The notation is as in App.~\ref{app:fkfpi}.


\vspace{1cm}

\begin{thebibliography}{10}

%\cite{Aoki:2013ldr}
\bibitem{Aoki:2013ldr}
  S.~Aoki {\it et al.}, 
  Review of lattice results concerning low energy particle physics,
  arXiv:1310.8555 [hep-lat].
  %%CITATION = ARXIV:1310.8555;%%
  %1 citations counted in INSPIRE as of 08 Nov 2013



%\cite{Bijnens:2006zp}
\bibitem{Bijnens:2006zp}
  J.~Bijnens,
  %``Chiral perturbation theory beyond one loop,''
  Prog.\ Part.\ Nucl.\ Phys.\  {\bf 58} (2007) 521
  [hep-ph/0604043].
  %%CITATION = HEP-PH/0604043;%%
  %104 citations counted in INSPIRE as of 27 Sep 2013



%\cite{Ecker:2010nc}
\bibitem{Ecker:2010nc}
  G.~Ecker, P.~Masjuan and H.~Neufeld,
  %``Chiral extrapolation and determination of low-energy constants
  %from lattice data,'' 
  Phys.\ Lett.\ B {\bf 692} (2010) 184
  [arXiv:1004.3422 [hep-ph]].
  %%CITATION = ARXIV:1004.3422;%%
  %4 citations counted in INSPIRE as of 27 Sep 2013



%\cite{Bijnens:1999hw}
\bibitem{Bijnens:1999hw}
  J.~Bijnens, G.~Colangelo and G.~Ecker,
  %``Renormalization of chiral perturbation theory to order $p^6$,''
  Annals Phys.\  {\bf 280} (2000) 100
  [hep-ph/9907333].
  %%CITATION = HEP-PH/9907333;%%
  %165 citations counted in INSPIRE as of 27 Sep 2013



%\cite{Gasser:1983yg}
\bibitem{Gasser:1983yg}
  J.~Gasser and H.~Leutwyler,
  %``Chiral Perturbation Theory to One Loop,''
  Annals Phys.\  {\bf 158} (1984) 142.
  %%CITATION = APNYA,158,142;%%
  %3071 citations counted in INSPIRE as of 27 Sep 2013



%\cite{Gasser:1984gg}
\bibitem{Gasser:1984gg}
  J.~Gasser and H.~Leutwyler,
  %``Chiral Perturbation Theory: Expansions in the Mass of the Strange Quark,''
  Nucl.\ Phys.\ B {\bf 250} (1985) 465.
  %%CITATION = NUPHA,B250,465;%%
  %2959 citations counted in INSPIRE as of 27 Sep 2013



%\cite{Bijnens:1999sh}
\bibitem{Bijnens:1999sh}
  J.~Bijnens, G.~Colangelo and G.~Ecker,
  %``The Mesonic chiral Lagrangian of order $p^6$,''
  JHEP {\bf 9902} (1999) 020
  [hep-ph/9902437].
  %%CITATION = HEP-PH/9902437;%%
  %257 citations counted in INSPIRE as of 27 Sep 2013



%\cite{Bijnens:1998yu}
\bibitem{Bijnens:1998yu}
  J.~Bijnens, G.~Colangelo and G.~Ecker,
  %``Double chiral logs,''
  Phys.\ Lett.\ B {\bf 441} (1998) 437
  [hep-ph/9808421].
  %%CITATION = HEP-PH/9808421;%%
  %47 citations counted in INSPIRE as of 27 Sep 2013


%\cite{Amoros:1999dp}
\bibitem{Amoros:1999dp}
  G.~Amor\'os, J.~Bijnens and P.~Talavera,
  %``Two point functions at two loops in three flavor chiral
  %perturbation theory,'' 
  Nucl.\ Phys.\ B {\bf 568} (2000) 319
  [hep-ph/9907264].
  %%CITATION = HEP-PH/9907264;%%
  %108 citations counted in INSPIRE as of 27 Sep 2013



%\cite{Bernard:2009ds}
\bibitem{Bernard:2009ds}
  V.~Bernard and E.~Passemar,
  %``Chiral Extrapolation of the Strangeness Changing $K \pi$ Form Factor,''
  JHEP {\bf 1004} (2010) 001
  [arXiv:0912.3792 [hep-ph]] and private communication.
  %%CITATION = ARXIV:0912.3792;%%
  %12 citations counted in INSPIRE as of 27 Sep 2013


%\cite{Durr:2010hr}
\bibitem{Durr:2010hr}
  S.~D{\"u}rr {\it et al.} [BMW Collaboration],
  %``The ratio $F_K/F_\pi$ in QCD,''
  Phys.\ Rev.\ D {\bf 81} (2010) 054507
  [arXiv:1001.4692 [hep-lat]].
  %%CITATION = ARXIV:1001.4692;%%
  %49 citations counted in INSPIRE as of 27 Sep 2013

%\cite{Amoros:2001cp}
\bibitem{Amoros:2001cp}
  G.~Amor\'os, J.~Bijnens and P.~Talavera,
  %QCD isospin breaking in meson masses, decay constants and quark mass ratios,
  Nucl.\ Phys.\ B {\bf 602} (2001) 87
  [hep-ph/0101127].
  %%CITATION = HEP-PH/0101127;%%
  %195 citations counted in INSPIRE as of 25 Dec 2013


%\cite{Bijnens:2011tb}
\bibitem{Bijnens:2011tb}
  J.~Bijnens and I.~Jemos,
  %``A new global fit of the $L^r_i$ at next-to-next-to-leading order
  %in Chiral Perturbation Theory,'' 
  Nucl.\ Phys.\ B {\bf 854} (2012) 631
  [arXiv:1103.5945 [hep-ph]].
  %%CITATION = ARXIV:1103.5945;%%
  %23 citations counted in INSPIRE as of 27 Sep 2013

\bibitem{Gasser:2007sg}
  J.~Gasser, C.~Haefeli, M.~A.~Ivanov and M.~Schmid,
  %Integrating out strange quarks in ChPT,
  Phys.\ Lett.\ B {\bf 652} (2007) 21
  [arXiv:0706.0955 [hep-ph]].
  %%CITATION = ARXIV:0706.0955;%%
  %40 citations counted in INSPIRE as of 14 Nov 2013



%\cite{Aoki:2010dy}
\bibitem{Aoki:2010dy}
  Y.~Aoki {\it et al.}  [RBC and UKQCD Collaborations],
  %``Continuum Limit Physics from 2+1 Flavor Domain Wall QCD,''
  Phys.\ Rev.\ D {\bf 83} (2011) 074508
  [arXiv:1011.0892 [hep-lat]].
  %%CITATION = ARXIV:1011.0892;%%
  %89 citations counted in INSPIRE as of 27 Sep 2013



%\cite{Arthur:2012opa}
\bibitem{Arthur:2012opa}
  R.~Arthur {\it et al.}  [RBC and UKQCD Collaborations],
  %``Domain Wall QCD with Near-Physical Pions,''
  Phys.\ Rev.\ D {\bf 87} (2013) 094514
  [arXiv:1208.4412 [hep-lat]].
  %%CITATION = ARXIV:1208.4412;%%
  %19 citations counted in INSPIRE as of 27 Sep 2013



%\cite{Beringer:1900zz}
\bibitem{Beringer:1900zz}
  J.~Beringer {\it et al.}  [Particle Data Group Collaboration],
  %``Review of Particle Physics (RPP),''
  Phys.\ Rev.\ D {\bf 86} (2012) 010001.
  %%CITATION = PHRVA,D86,010001;%%
  %2259 citations counted in INSPIRE as of 27 Sep 2013


%\cite{DescotesGenon:1999uh}
\bibitem{DescotesGenon:1999uh}
  S.~Descotes-Genon, L.~Girlanda and J.~Stern,
  %``Paramagnetic effect of light quark loops on chiral symmetry breaking,''
  JHEP {\bf 0001} (2000) 041
  [hep-ph/9910537].
  %%CITATION = HEP-PH/9910537;%%
  %81 citations counted in INSPIRE as of 27 Sep 2013



%\cite{Bazavov:2009bb}
\bibitem{Bazavov:2009bb}
  A.~Bazavov {\it et al.},
  %``Nonperturbative QCD simulations with 2+1 flavors of improved
  %staggered quarks,'' 
  Rev.\ Mod.\ Phys.\  {\bf 82} (2010) 1349
  [arXiv:0903.3598 [hep-lat]].
  %%CITATION = ARXIV:0903.3598;%%
  %209 citations counted in INSPIRE as of 27 Sep 2013



%\cite{Ademollo:1964sr}
\bibitem{Ademollo:1964sr}
  M.~Ademollo and R.~Gatto,
  %``Nonrenormalization Theorem for the Strangeness Violating Vector
  %Currents,'' 
  Phys.\ Rev.\ Lett.\  {\bf 13} (1964) 264.
  %%CITATION = PRLTA,13,264;%%
  %425 citations counted in INSPIRE as of 27 Sep 2013



%\cite{Gasser:1984ux}
\bibitem{Gasser:1984ux}
  J.~Gasser and H.~Leutwyler,
  %``Low-Energy Expansion of Meson Form-Factors,''
  Nucl.\ Phys.\ B {\bf 250} (1985) 517.
  %%CITATION = NUPHA,B250,517;%%
  %591 citations counted in INSPIRE as of 27 Sep 2013



%\cite{Bazavov:2012cd}
\bibitem{Bazavov:2012cd}
  A.~Bazavov {\it et al.} [Fermilab Lattice and MILC Collaborations],
  %``Kaon semileptonic vector form factor and determination of
  %$|V_{us}|$ using staggered fermions,'' 
  Phys.\ Rev.\ D {\bf 87} (2013) 073012
  [arXiv:1212.4993 [hep-lat]].
  %%CITATION = ARXIV:1212.4993;%%
  %9 citations counted in INSPIRE as of 27 Sep 2013



%\cite{Boyle:2013gsa}
\bibitem{Boyle:2013gsa}
  P.~A.~Boyle {\it et al.}  [RBC and UKQCD Collaborations], 
  %``The kaon semileptonic form factor with near physical domain wall quarks,''
  JHEP {\bf 1308} (2013) 132
  [arXiv:1305.7217 [hep-lat]].
  %%CITATION = ARXIV:1305.7217;%%
  %1 citations counted in INSPIRE as of 27 Sep 2013



%\cite{Post:2001si}
\bibitem{Post:2001si}
  P.~Post and K.~Schilcher,
  %``$K_{l3}$ form-factors at order $p^6$ of chiral perturbation theory,''
  Eur.\ Phys.\ J.\ C {\bf 25} (2002) 427
  [hep-ph/0112352].
  %%CITATION = HEP-PH/0112352;%%
  %72 citations counted in INSPIRE as of 27 Sep 2013



%\cite{Bijnens:2003uy}
\bibitem{Bijnens:2003uy}
  J.~Bijnens and P.~Talavera,
  %``K(l3) decays in chiral perturbation theory,''
  Nucl.\ Phys.\ B {\bf 669} (2003) 341
  [hep-ph/0303103].
  %%CITATION = HEP-PH/0303103;%%
  %164 citations counted in INSPIRE as of 27 Sep 2013



%\cite{Weinberg:1978kz}
\bibitem{Weinberg:1978kz}
  S.~Weinberg,
  %``Phenomenological Lagrangians,''
  Physica A {\bf 96} (1979) 327.
  %%CITATION = PHYSA,A96,327;%%
  %2384 citations counted in INSPIRE as of 27 Sep 2013

\end{thebibliography}
\end{document}